\documentclass[twocolumn,aps,pre,floatfix,superscriptaddress]{revtex4}
\pdfoutput=1 
\usepackage{amsmath,amssymb,eucal}
\usepackage{graphicx}

\begin{document}
\title{Assortative Exchange Processes}

\author{P.~L.~Krapivsky}
\affiliation{Department of Physics, Boston University, Boston, MA 02215, USA}

\begin{abstract}
In exchange processes clusters composed of elementary building blocks, monomers, undergo binary exchange in which a monomer is transferred from one cluster to another. In assortative exchange only clusters with comparable masses participate in exchange events. We study maximally assortative exchange processes in which only clusters of equal masses can exchange monomers. A mean-field framework based on rate equations is appropriate for spatially homogeneous systems in sufficiently high spatial dimension. For diffusion-controlled exchange processes, the mean-field approach is erroneous when the spatial dimension is smaller than critical; we analyze such systems using scaling and heuristic arguments. Apart from infinite-cluster systems we explore the fate of finite systems and study maximally assortative exchange processes driven by a localized input. 
\end{abstract}
\maketitle

\maketitle

\section{Introduction}

Exchange processes arise in numerous natural phenomena such as droplet growth via evaporation and re-condensation \cite{meakin}, island growth \cite{az} and phase ordering \cite{ls,ajb,sm}.  Exchange processes have been applied to social sciences, e.g., to modeling segregation of heterogeneous populations \cite{ts}, studying the distribution of wealth through asset exchange \cite{Angle86,IKR98,CC00,Hayes,Angle06,Boghosian15}, mimicking growth of urban populations \cite{lr} and aggregation behaviors in job markets \cite{Sun14}. Exchange processes are also used as toy microscopic models which are simple enough to allow the derivation of the macroscopic `hydrodynamic' equations and explore other fundamental aspects of non-equilibrium statistical mechanics, see e.g. \cite{GG09,Khanin12,Evans12,Colm15,GG17} and references therein.

In mass exchange processes, clusters interact by transferring mass from one to another. Cluster are usually assumed to be composed of an integer number of elemental building blocks (`monomers'). We shall denote by $A_j$ a cluster of `mass' $j$, that is, a cluster which is made of $j$ monomers. Clusters are thus labelled solely by their masses; other characteristics (e.g., their shape) are ignored. We assume that in each exchange event, a monomer is transferred from one cluster to another.  Symbolically, the mass exchange process is represented by the reaction scheme 
\begin{equation}
\label{Aij}
A_i \oplus A_j\mathop{\longrightarrow}^{K_{i,j}} \left(A_{i\pm 1}, A_{j\mp 1}\right)
\end{equation}
A cluster disappears when its mass vanishes. Thus in an exchange involving a monomer the number of clusters may decrease; it certainly decreases in an exchange between monomers, one monomer disappears and another becomes a dimer. 

Exchange processes characterized by symmetric migration rates $K_{i,j}$ have been mostly investigated, e.g., models with homogeneous rates $K_{i,j}=i^aj^b+i^bj^a$ have been studied through asymptotic and scaling analyses (see e.g. \cite{KL02,EP03,KL06}). Even in the simplest situation when the system is spatially uniform and remains well-mixed throughout the evolution, the governing (mean-field) rate equations form an infinite set of coupled non-linear differential equations which could not be solved. The exchange processes characterized by the generalized product kernel $K_{i,j}=(ij)^a$ are special since the governing equations can be linearized, and the models with $a=0,1,2$ have been solved exactly, see \cite{EP03,KL06,book,PK_exchange}. 

In assortative exchange processes, interactions between clusters with disparate masses is suppressed. Here we consider the maximally assortative processes in which exchange can occur only between clusters of the same mass. The matrix of migration rates becomes diagonal, $K_{i,j}=K(i) \delta_{i,j}$, and the set of reaction channels \eqref{Aij} narrows to
\begin{equation}
\label{Amm}
A_m \oplus A_m\mathop{\longrightarrow}^{K(m)} \left(A_{m-1}, A_{m+1}\right)
\end{equation}

We now outline the mathematical framework and announced a few chief results. The rate equations governing the evolution of the general maximally assortative exchange process are 
\begin{equation}
\label{exchange:assort}
\frac{dc_m}{d t}=K(m+1) c_{m+1}^2-2K(m) c_m^2+K(m-1) c_{m-1}^2
\end{equation}
Here $c_m(t)$ is the density  of clusters containing $m$ monomers at time $t$. The density of monomers obeys $\dot c_1 = K(2)c_2^2-2K(1)c_1^2$ which is consistent with the first equation  \eqref{exchange:assort} after setting $c_0\equiv 0$, or introducing an extra rate $K(0)=0$. One can verify that Eqs.~\eqref{exchange:assort} agree with mass conservation:
\begin{equation}
\label{mass}
\sum_{m\geq 1} mc_m(t) = 1
\end{equation}
Hereinafter we set the conserved mass density $M$ to unity; this can always be done by rescaling: $c_m\to Mc_m$.

Equations \eqref{exchange:assort} have not been solved; the only exception are somewhat pathological models in which for a certain mass $j$ the corresponding rate vanishes, $K(j)=0$, so only clusters up to mass $j$ are present. In the following we ignore such models and assume that $K(m)>0$ for all $m\geq 1$, so that the number of interacting cluster species is infinite. The most interesting long time behavior of such models can be probed through asymptotic and scaling analyses. For instance, when rates are mass-independent, the density of monomers decays according to
\begin{equation}
\label{c1t}
c_1 \sim
\begin{cases}
t^{-5/8}                     & d=3\\
t^{-5/8} (\ln t)^{5/8}   & d=2\\
t^{-7/18}                    & d=1
\end{cases}
\end{equation}

To appreciate these results we first recall that Eqs.~\eqref{exchange:assort} are applicable only if clusters remain perfectly mixed throughout evolution. The analysis of Eqs.~\eqref{exchange:assort} with kernel $K(m)=1$ leads to $c_1\sim t^{-5/8}$ as we show in Sect.~\ref{sec:const}. Now one can ask about actual physical process which could be mathematically described by Eqs.~\eqref{exchange:assort} with kernel $K(m)=1$. In the case of diffusive transport, the natural candidate is the point cluster process in which
\begin{enumerate}
\item Each clusters occupies a single lattice site of a $d-$dimensional lattice. 
\item Clusters hop to neighboring sites and  hopping rates are mass-independent. 
\item When a cluster hops to a site containing a cluster with the same mass, an exchange \eqref{Amm} instantaneously occurs. 
\end{enumerate}

A critical dimension for this diffusion-controlled exchange process is $d_c =2$, that is the rate equations Eqs.~\eqref{exchange:assort} with mass-independent kernel describe the evolution when $d>d_c=2$, particularly in three dimensions. At the critical dimension there is a logarithmic correction to the mean-field behavior; below the critical dimension the decay is slower than the mean-field prediction (similarly to other diffusion-controlled processes, see \cite{book}). It is often useful to treat $d$ as a continuous parameter. The decay exponents are universal when $d>d_c=2$ and become dimension-dependent when $d<d_c$ where, as we argue in Sect.~\ref{sec:d}, the density of monomers decays as
\begin{equation}
\label{c1:below}
\frac{c_1}{n_0}\sim \left[n_0 (Dt)^{d/2}\right]^{-\frac{3d+4}{(d+2)^2}}
\end{equation}
Here we write the answer in the dimensionally-correct form which demonstrates the dependence on the diffusion coefficient $D$ and the initial density $n_0$; the latter is defined via $c_m(0)=n_0\delta_{m,1}$ if the system is initially composed of monomers. 

Further, the cluster density 
\begin{equation}
\label{density}
N(t) = \sum_{m\geq 1} c_m(t)
\end{equation}
decays according to
\begin{equation}
\label{Nt}
N \sim
\begin{cases}
t^{-1/4}                     & d=3\\
t^{-1/4} (\ln t)^{1/4}   & d=2\\
t^{-1/6}                     & d=1
\end{cases}
\end{equation}
More generally, below the critical dimension, $d<2$, the cluster density decays as 
\begin{equation}
\label{N:below}
\frac{N}{n_0}\sim \left[n_0 (Dt)^{d/2}\right]^{-\frac{1}{d+2}}
\end{equation}

In Sect.~\ref{sec:const} we study the asymptotic behavior of the solutions to Eqs.~\eqref{exchange:assort} with mass-independent migration rates, $K(m)=1$. In Sect.~\ref{sec:alg} we extend results of Sect.~\ref{sec:const} to a one-parameter family of rates varying algebraically with mass, $K(m)=m^a$. Such rates are particularly suitable for scaling techniques which we employ. The analysis of Sects.~\ref{sec:const} and \ref{sec:alg} relies on mean-field equations \eqref{exchange:assort} which are valid if the initial state is spatially homogeneous and if the system remains well-mixed throughout the evolution. In Sect.~\ref{sec:d} we discuss the behavior of the simplest diffusion-controlled maximally assortative exchange process in which each cluster occupies a single lattice site (the point cluster process) and hops with mass-independent rate. The critical dimension is $d_c=2$ for such exchange processes, and we analyze asymptotic behaviors of these processes in one and two dimensions. Maximally assortative exchange processes with a finite mass generically do not condense in a single cluster, but reach a non-trivial final state with numerous clusters with different masses. In Sect.~\ref{sec:FS} we describe these final states. Diffusion-controlled maximally assortative exchange processes driven by a localized input of monomers are investigated in Sect.~\ref{sec:input}. In Sect.~\ref{sec:disc} we discuss approaches which may lead to the progress in understanding of maximally assortative exchange processes with quickly growing rates where scaling is violated.

\section{Mass-Independent Rates}
\label{sec:const}

For the maximally assortative exchange process with mass-independent rates, $K(m)=1$, Eqs.~\eqref{exchange:assort} reduce to
\begin{equation}
\label{exchange:const}
\frac{dc_m}{d t} = c_{m+1}^2 - 2 c_m^2 + c_{m-1}^2
\end{equation}
This neat infinite system of non-linear coupled ordinary differential equations (ODEs) appears intractable. The most interesting large time behavior can be established, however, since the typical mass exhibits an unlimited growth when $t\to \infty$ thereby allowing us to employ asymptotic and scaling approaches. The chief idea is to treat $m$ as a continuous variable. If $c(m,t)\equiv c_m(t)$ slowly varies with $m$, the right-hand side in \eqref{exchange:const} can be replaced by the second derivative to yield
\begin{equation}
\label{exchange:PDE}
\frac{\partial c}{\partial t} = \frac{\partial^2 }{\partial m^2}\,c^2
\end{equation}
One then seeks a scaling solution to \eqref{exchange:PDE}:
\begin{equation}
\label{scaling}
c(m,t) = t^{-2\beta} F(x), \qquad x=\frac{m}{t^\beta}
\end{equation}
The scaling form \eqref{scaling} agrees with mass conservation, the choice \eqref{mass} of the initial mass density implies
\begin{equation}
\label{mass:F}
\int_0^\infty dx\,xF(x) = 1
\end{equation}
By inserting the scaling form \eqref{scaling} into \eqref{exchange:PDE} we find that the scaling form is consistent only when $\beta=1/4$, and in that case the governing PDE turns into an ODE
\begin{equation}
\label{exchange:ODE}
4(F^2)'' + xF' + 2F = 0
\end{equation}
where $(\cdot)' = d(\cdot)/dx$. Multiplying \eqref{exchange:ODE} by $x$ and integrating we get
\begin{equation}
\label{F1:eq}
4\,\frac{d}{dx}\left(\frac{F^2}{x}\right) + F = 0
\end{equation}
where the integration constant was chosen to be zero to assure that $F$ vanishes as $x\to\infty$. Re-writing Eq.~\eqref{F1:eq} as a product of two factors, $F[8F'/x+1-4F/x^2]=0$, we immediately extract the special solution $F(x)=0$, and then from $8F'/x+1-4F/x^2=0$ we find a one-parameter family of solutions
\begin{equation}
\label{F0}
F = \frac{1}{12}\,\sqrt{x}\,\big(x_0^{3/2}-x^{3/2}\big)
\end{equation}

The solution is the combination of \eqref{F0} and $F=0$. Since the density is non-negative, the scaled mass distribution is given by \eqref{F0} when $0\leq x\leq x_0$ and $F(x)=0$ for $x>x_0$.  The parameter $x_0$ is found from the normalization requirement, $1=\int_0^{x_0} dx\,xF(x)$, yielding $x_0=(80)^{1/4}$. Hereinafter it proves convenient to use a renormalized scaled mass variable $y = x/x_0$ which in the present case equals to $y=m/(80 t)^{1/4}$ in terms of the original variables. Collecting previous results we write the scaling solution in the form
\begin{equation}
\label{cmt:0}
c_m(t) = t^{-1/2}\,G(y)
\end{equation}
with 
\begin{equation}
\label{G:0}
G(y) = \frac{1}{3}\times 
\begin{cases}
\sqrt{5y}\,\big(1-y^{3/2}\big)  & 0\leq y\leq 1\\
0                                         & y>1
\end{cases}
\end{equation}
Note also simple asymptotic formulas 
\begin{subequations}
\begin{align}
\label{c1:0}
c_1(t) &= B_1\, t^{-5/8}\\
\label{N:0}
N(t) &= B\,  t^{-1/4}
\end{align}
\end{subequations}
for the density of monomers and the total cluster density $N(t)=\sum_{m\geq 1}c_m(t)$. The amplitudes in \eqref{c1:0}--\eqref{N:0} are 
\begin{equation*}
B_1 = \frac{5^{3/8}}{3\sqrt{2}}\,, \qquad B = \frac{2\cdot 5^{3/4}}{9}
\end{equation*}

At first sight, the compact shape of the mass distribution seems paradoxical given that the governing equations are parabolic PDEs. Our intuition is based on linear parabolic PDEs for which perturbation propagates with infinite speed preventing the formation of compact solutions. For {\em non-linear} parabolic PDEs like Eqs.~\eqref{exchange:PDEa}, however, compact solutions may and do arise as was discovered many years ago \cite{ZK,B52}, see \cite{Zeld,LL87,B96} for review and \cite{Landim:09,Pablo12} for recent examples of such non-linear parabolic PDEs appearing in the context of lattice gases. 

The compactness is a drawback of the continuum approximation. In the realm of the original discrete system \eqref{exchange:assort}, the mass distribution is positive for all $m$. The front is extremely steep, however, so the discrepancy between the actual solution of the discrete system and the prediction of the continuum approach, viz. $c(m,t)=0$ in the region $m> m_*(t)=(80 t)^{1/4}$, is tiny. To appreciate this we take into account a very sharp decay and simplify \eqref{exchange:const} in the $m> m_*(t)$ region to
$\frac{dc_m}{d t} \simeq c_{m-1}^2$ from which we deduce a double exponential decay
\begin{equation}
\ln(1/c_m)\propto 2^{m-m_*(t)}
\end{equation}

\section{Arbitrary Homogeneous Rates}
\label{sec:alg}

In this section we investigate maximally assortative exchange processes with algebraically varying migration rates $K(m)=m^a$. The governing rate equations read 
\begin{equation}
\label{exchange:a}
\frac{dc_m}{d t}=(m+1)^a c_{m+1}^2-2m^a c_m^2+(m-1)^a c_{m-1}^2
\end{equation}
Note that a rate equation for the cluster density
\begin{equation}
\label{N:gen}
\frac{dN}{d t}= - c_1^2
\end{equation}
is independent on the exponent $a$. 

\subsection{Scaling approach}
\label{sec:scaling}

Algebraically varying migration rates are physically natural and convenient for analysis since they are compatible with the scaling approach. Thus we immediately focus on the large time behavior, treat again $m$ as a continuous variable and turn an infinite set of ODEs, Eqs.~\eqref{exchange:a}, into a single PDE
\begin{equation}
\label{exchange:PDEa}
\frac{\partial c}{\partial t} = \frac{\partial^2 }{\partial m^2}\,m^a c^2
\end{equation}
The scaling solution to this PDE has the form \eqref{scaling} with $\beta=(4-a)^{-1}$; the scaled mass distribution obeys
\begin{equation}
\label{exchange:ODEa}
(4-a) (x^a F^2)'' + xF' + 2F = 0
\end{equation}
Multiplying \eqref{exchange:ODEa} by $x$ and integrating once we obtain
\begin{equation}
(4-a)\,\frac{d}{dx}\left(\frac{F^2}{x^{1-a}}\right) + F = 0
\end{equation}
This equation admits the special solution $F(x)=0$ and a one-parameter family of solutions
\begin{equation}
\label{Fa}
G = \frac{x_0^{2-a}}{(3-a)(4-a)}\,y^{\frac{1-a}{2}}\left(1-y^{\frac{3-a}{2}}\right)
\end{equation}
where we have used again the renormalized scaled mass variable $y=x/x_0$. The scaled mass distribution is given by \eqref{Fa} when $0\leq y\leq 1$, while $G(y)=0$ for $y>1$. The parameter $x_0$ is fixed by normalization:
\begin{equation}
\label{x0:a}
x_0 = (4-a)^{2/(4-a)} (5-a)^{1/(4-a)}
\end{equation}

Gathering previous results we arrive at 
\begin{equation}
\label{cmt:a}
\begin{split}
& c_m(t) = t^{-\frac{2}{4-a}}\,G_a(y) \\
& y = \frac{x}{x_0} = \left[(4-a)^2(5-a)t\right]^{-\frac{1}{4-a}}\,m
\end{split}
\end{equation}
The scaled mass distribution reads 
\begin{equation}
\label{Ga}
G_a(y) = C(a)\times 
\begin{cases}
y^{\frac{1-a}{2}}\left(1-y^{\frac{3-a}{2}}\right)  & 0\leq y\leq 1\\
0                                                                     & y>1
\end{cases}
\end{equation}
The amplitude is
\begin{equation}
C(a) = \frac{x_0^{2-a}}{(3-a)(4-a)}
\end{equation}
with $x_0(a)$ given by \eqref{x0:a}. 

The density of monomers and the total cluster density exhibit algebraic long time behaviors
\begin{equation}
\label{c1c:a}
c_1 = B_1(a)\, t^{-(5-a)/(8-2a)}, \quad N = B(a)\,  t^{-1/(4-a)} 
\end{equation}
with amplitudes 
\begin{equation}
\label{BBa}
B_1(a) = \frac{x_0^{(3-a)/2}}{(3-a)(4-a)}\,, ~~ B(a) = \frac{x_0^{3-a}}{(3-a)^2 (4-a)}
\end{equation}

\begin{figure}
\centering
\includegraphics[width=7.77cm]{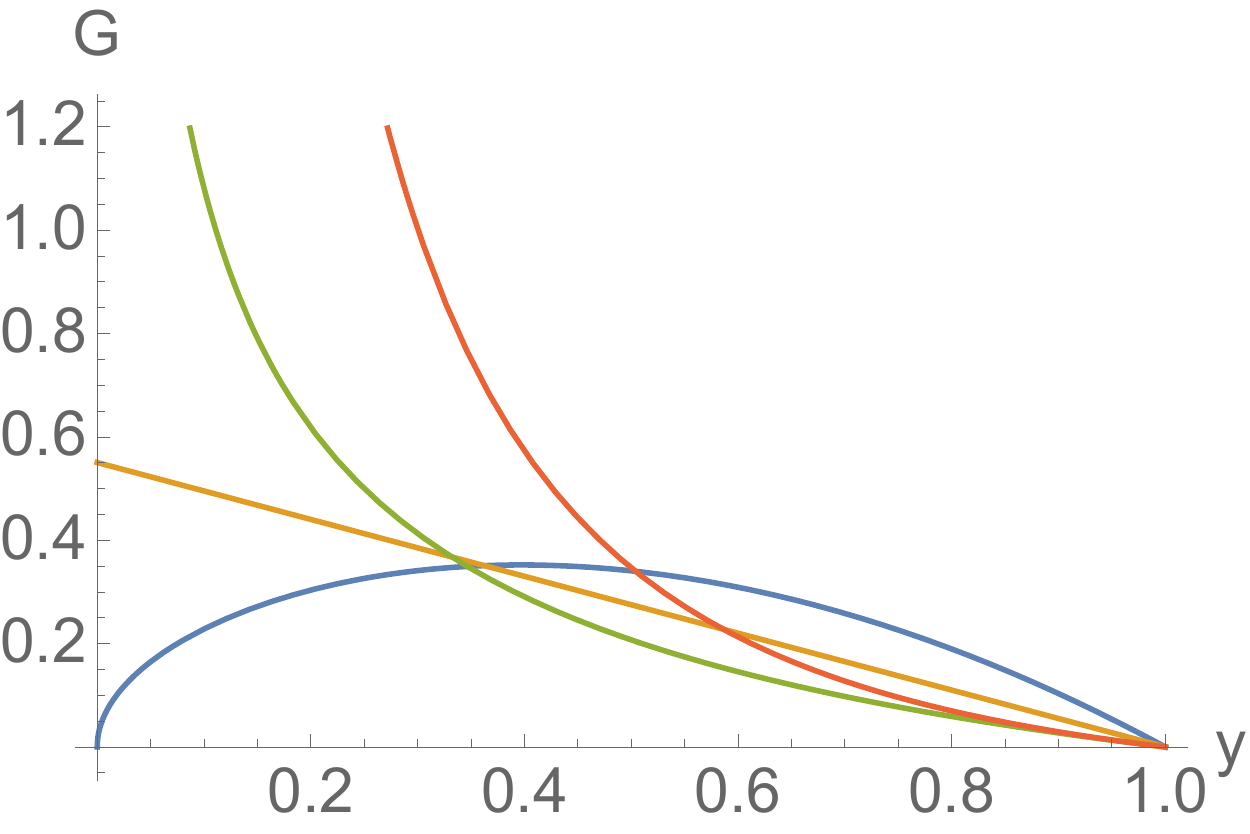}
\caption{Shown are the scaled mass distributions $G_a$, Eq.~\eqref{G0123}, for $a=0,1,2,3$ (top to bottom near $y=1$).}
\label{Fig:MD0123}
\end{figure}

On Fig.~\ref{Fig:MD0123} we plot the positive part of the scaled mass distribution for $a=0,1,2,3$:
\begin{equation}
\label{G0123}
\begin{split}
G_0 &= \sqrt{5y}\,\big(1-y^{3/2}\big)/3\\
G_1 &= 6^{-1/3}(1-y)\\
G_2 &= \big(y^{-1/2}-1\big)/2\\
G_3 &= (4y)^{-1}\ln(1/y)
\end{split}
\end{equation}

\subsection{Marginal case of $a=3$}

The results of Sect.~\ref{sec:scaling} are applicable only when the homogeneity index satisfies $a<3$. Taking the $a\uparrow 3$ limit in \eqref {Ga} we obtain consistent results, namely the scaled mass distribution becomes 
\begin{equation}
\label{cm3:scaling}
c_m(t) = t^{-2} G_3(y), \qquad y=\frac{m}{2t}
\end{equation}
with
\begin{equation}
\label{G3}
G_3(y) = 
\begin{cases}
(4y)^{-1}\ln(1/y)  & 0<y<1\\
0                         & y\geq 1
\end{cases}
\end{equation}
Specializing \eqref{cm3:scaling}--\eqref{G3} to $m=1$ we obtain 
\begin{subequations}
\begin{equation}
\label{c1:a3}
c_1\simeq \frac{\ln t}{2t}
\end{equation}
The decay law for the total cluster density also acquires a logarithmic correction
\begin{equation}
\label{N:a3}
N \simeq \frac{(\ln t)^2}{4t}
\end{equation}
\end{subequations}
To establish this decay law we use \eqref{cm3:scaling}--\eqref{G3} and find
\begin{equation}
\label{N:der}
N = \sum_{m\geq 1} mc_m(t) \simeq  \frac{2t}{t^2}\int_{1/(2t)}^1dy\,\frac{\ln(1/y)}{4y}
\end{equation}
Computing the integral yields \eqref{N:a3} in the leading order. Note that the integral in \eqref{N:der} diverges in the $y\to 0$ forcing us to keep the lower limit finite. This makes the replacement of the summation by integration somewhat doubtful, but it actually does not cause the problem since the divergence is logarithmic and hence the prediction should be correct. As an independent check we can use the exact rate equation \eqref{N:gen} and verify that it is consistent with \eqref{c1:a3}--\eqref{N:a3}.

\subsection{Non-scaling regime: $a>3$}

The scaling solution \eqref{cmt:a}--\eqref{Ga} does not make sense when $a>3$. The behavior in this region is non-scaling, so it is harder to probe analytically. Analogously to ordinary exchange processes, one may expect gelation. For instance, in ordinary exchange processes with generalized product kernels $K_{i,j}=(ij)^\lambda$, it has been shown \cite{EP03} that (i) scaling holds when $\lambda\leq \frac{3}{2}$; (ii) an infinite cluster (`gel') is formed at a finite time if $\frac{3}{2}<\lambda \leq 2$; (iii) a gel is formed at time  $t=+0$ when $\lambda>2$ and gelation is complete, i.e. $c_m(t)=0$ for all $m\geq 1$ at $t>0$. 

Below in Sect.\ref{sec:FS-extreme} we consider the maximally assortative exchange process with $a=\infty$. In this extremal model $K(m)=\infty$ for all $m\geq 2$ and the the emerging mass distribution has certain features resembling instantaneous gelation, yet there is no gel. More precisely, in the infinite-system limit the rate equations are mathematically ill-defined for the extremal model. We thus consider the extremal model with finite total mass $\mathcal{M}$ and establish [see \eqref{N1:inf}--\eqref{Nm:inf}] the following mass distribution
\begin{equation}
c_m =
\begin{cases} 
(1+t)^{-1}               & m=1\\
\mathcal{M}^{-1}   & 2\leq m\leq \sqrt{2\mathcal{M}\frac{t}{1+t}}
\end{cases}
\end{equation}
Thus in the $\mathcal{M}\to\infty$ limit all cluster densities apart from the monomer density vanish: $c_m(t)=0$ for all $m\geq 2$ at $t>0$. This is similar to instantaneous gelation. There is no gel, however. 

The knowledge of the behavior in the $a\leq 3$ range and the behavior at $a=\infty$ allows one to make rather plausible guesses about the behavior in the  $a>3$ range. For instance, the cluster density decays as 
\begin{equation}
\label{N:a}
N \simeq 
\begin{cases}
B(a)\,t^{-1/(4-a)}     & a<3\\
(4t)^{-1}(\ln t)^2      & a=3\\
t^{-1}                      & a=\infty
\end{cases}
\end{equation}
with $B(a)$ appearing in \eqref{BBa}. Thus the upper and lower bounds for $N(t)$ are
\begin{equation}
(4t)^{-1}(\ln t)^2 < N(t) < t^{-1}
\end{equation}
when $a>3$. Logarithmic corrections usually appear in the marginal cases, like $a=3$ in our situation, so for all $a>3$ we anticipate a simple decay
\begin{equation}
N \simeq  \frac{B^+(a)}{t}
\end{equation}
The unknown amplitude $B^+(a)$ should decrease from $\lim_{a\to 3+0}B^+(a)=\infty$ to $\lim_{a\to \infty}B^+(a)=1$. 

Similarly for the density of monomers we have established the following decay laws:
\begin{equation}
\label{c1:a}
c_1 \simeq 
\begin{cases}
B_1(a)\,t^{-1/(4-a)}     & a<3\\
(2t)^{-1} \ln t               & a=3\\
t^{-1}                         & a=\infty
\end{cases}
\end{equation}
with $B_1(a)$ appearing in \eqref{BBa}. Thus the bounds are 
\begin{equation}
(2t)^{-1} \ln t < c_1(t) < t^{-1}
\end{equation}
for $a>3$, and we actually expect a simple decay
\begin{equation}
c_1 \simeq  \frac{B_1^+(a)}{t}
\end{equation}
Similarly to $B^+(a)$, the amplitude $B^+_1(a)$ should decrease from $\lim_{a\to 3+0}B^+_1(a)=\infty$ to $\lim_{a\to \infty}B^+_1(a)=1$, and the inequality $B^+_1(a)\leq B^+(a)$ should be valid.

\section{Exchange processes in low spatial dimensions}
\label{sec:d}

Here we probe the behavior of diffusion-controlled exchange processes. We assume that clusters hop on a lattice and each cluster occupies a single lattice. The number of clusters at a site is unlimited, but all such clusters are such that they cannot mutually participate in an exchange. When a cluster hops to a site that contains another cluster which can participate in an exchange process, the exchange instantaneously occurs; if then another exchange becomes possible it also occurs instantaneously. Overall, an avalanche of exchanges may happen. 

If exchange between clusters of arbitrary masses is allowed, the diffusion-controlled exchange process is actually identical to the diffusion-controlled aggregation process: When a cluster $A_i$ hops to a site with cluster $A_j$, an avalanche of exchanges occur till eventually a single cluster $A_{i+j}$ is formed, this happens instantaneously so the process is indeed a merging event $A_i \oplus A_j\to A_{i+j}$. We shall always assume that the hopping rates are mass-independent. The diffusion-controlled aggregation process in which clusters occupy single sites and hop with mass-independent rates is well-understood, the critical dimension is known to be $d_c=2$; the computations of the critical dimension in various diffusion-controlled processes are described e.g. \cite{vD89} and \cite{book}. The mean-field framework reproduces the asymptotic behavior above the critical dimension, $d>d_c=2$, e.g. it correctly predicts the $N\sim t^{-1}$ decay of the cluster density. In two dimensions, the mean-field framework is almost correct as it only misses logarithmic factors, e.g. $N\sim t^{-1}\ln t$. In one dimension, the deviations from the mean-field behavior are most pronounced, e.g. $N\sim t^{-1/2}$.  The one-dimensional diffusion-controlled aggregation process is actually solvable, see \cite{S88,TNT88,T89}. 

Let us now consider the maximal assortative exchange in which a monomer can be transferred only between clusters with equal masses: $A_m \oplus A_m \to \left(A_{m-1}, A_{m+1}\right)$. In principle, an avalanche of exchanges can occur. As an example, consider what happens when $A_2$ hops to a site containing clusters $(A_1, A_2, A_3)$. After the hop there are four clusters $(A_1, A_2, A_2, A_3)$ at a site and an avalanche of exchanges (participating pairs  are shown) leads to 
\begin{equation*}
A_1\underbrace{A_2 A_2} A_3 \to \underbrace{A_1 A_1} \underbrace{A_3 A_3}
\to \underbrace{A_2 A_2} A_4 \to A_1 A_3 A_4
\end{equation*}
Thus effectively $A_2\oplus (A_1, A_2, A_3)\to (A_1, A_3, A_4)$. 

In the long time limit the density of clusters approaches to zero and avalanches become exceedingly rare. Indeed, the exchange $A_m \oplus A_m \to \left(A_{m-1}, A_{m+1}\right)$ results in two clusters at a site, but these clusters quickly separate so that when $t\gg 1$ an occupied lattice site is almost surely occupied by a single clusters.

This problem appears analytically intractable even in one dimension, so we focus on the simplest characteristics, the decay exponents, and rely on heuristic arguments. As a check of such arguments let us first recover the exponents describing the decay of the monomer and cluster densities which we know from the asymptotically exact analysis, see \eqref{c1:0}--\eqref{N:0}. 

In the long time limit the left-hand side in \eqref{exchange:const} decays faster than the terms in the right-hand side, so the right-hand side must asymptotically vanish. We thus get $c_{m+1}^2 - 2 c_m^2 + c_{m-1}^2=0$. The general solution is $c_m^2=A m + B$; recalling the convention $c_0=0$ we conclude that $B=0$. Using $c_m\propto \sqrt{m}$ in conjunction with the scaling form \eqref{scaling} we obtain $F(x)\sim \sqrt{x}$ and then $c_m\sim \sqrt{m/t^{5\beta}}$ when $m\ll t^\beta$.   Combining $c_1\sim t^{-5\beta/2}$ and $N\sim t^{-\beta}$ with the exact rate equation
\begin{equation}
\label{Nc}
\frac{dN}{dt} = - c_1^2
\end{equation}
we find $\beta=1/4$ and therefore
\begin{equation}
\label{c1N}
c_1\sim t^{-5/8}\,, \qquad N\sim t^{-1/4}
\end{equation}
recovering the exponents in \eqref{c1:0}--\eqref{N:0}.  

Consider now the maximal assortative exchange on the one-dimensional lattice. (In one dimension the lattice version is not necessary, we can treat clusters as point particles performing independent Brownian motions with the same diffusion coefficient $D$.) To estimate the decay of the total density, consider two adjacent monomers. They are separated by distance $\ell\sim c_1^{-1}$, and it takes time $T\sim \ell^2/D\sim D^{-1}c_1^{-2}$ for these monomers to meet. Thus the cluster density decays according to
\begin{equation}
\label{Nc:1d}
\frac{dN}{dt} \sim - \frac{c_1}{T} \sim -Dc_1^3
\end{equation}
We now use rate equations similar to \eqref{exchange:const}, but with $c_m^3$ instead of $c_m^2$ in the right-hand side. (This step involves an uncontrolled approximation, but such approximations have been used in various diffusion-controlled processes and they always lead to correct asymptotic behaviors.) In the long time limit we thus obtain $c_{m+1}^3 - 2 c_m^3 + c_{m-1}^3=0$, from which $c_m\propto \sqrt[3]{m}$. Combining with the scaling form \eqref{scaling} we obtain  $c_m\sim \sqrt[3]{m/t^{7\beta}}$ when $m\ll t^\beta$. Plugging $c_1\sim t^{-7\beta/3}$ and $N\sim t^{-\beta}$ into \eqref{Nc:1d} we deduce $\beta=1/6$. Thus
\begin{equation}
\label{c1N:1d}
c_1\sim t^{-7/18}\,, \qquad N\sim t^{-1/6}
\end{equation}
In the dimensionally correct form the decay laws read 
\begin{equation}
\label{c1N:1d-full}
c_1\sim n_0^{2/9}(Dt)^{-7/18}\,, \quad N\sim n_0^{2/3}(Dt)^{-1/6}
\end{equation}
where $n_0$ is the initial density of monomers. 

In two spatial dimensions, we similarly get
\begin{equation}
\label{Nc:2d}
\frac{dN}{dt}  \sim  \frac{Dc_1^2}{\ln[c_1a^2]}
\end{equation}
where $a$ is the lattice spacing. The same argument as before leads to $c_{m+1}^2 - 2 c_m^2 + c_{m-1}^2=0$; more precisely one should write $c_m^2/\ln[c_m a^2]$, but since $c_m$ decay in time according to the same law, the logarithmic factor is independent on $m$ in the leading order. Thus $c_m\propto \sqrt{m}$ and more precisely $c_m\sim \sqrt{m/\mu^{5}}$, where we use
\begin{equation}
\label{scaling:mu}
c(m,t) = \mu^{-2} F(x), \qquad x=\frac{m}{\mu}
\end{equation}
We denote the typical size by $\mu$ rather than $t^\beta$ since in addition to the algebraic factor $\mu$ has a logarithmic factor. Plugging $c_1\sim \mu^{-5/2}$ and $N\sim \mu^{-1}$ into \eqref{Nc:2d} we find $\mu\sim (t/\ln t)^{1/4}$ leading to
\begin{equation}
\label{c1N:2d}
c_1\sim \left(\frac{\ln t}{t}\right)^{5/8}\,, \quad N\sim \left(\frac{\ln t}{t}\right)^{1/4}
\end{equation}
In the dimensionally correct form the decay laws read 
\begin{equation}
\label{c1N:2d-full}
c_1\sim n_0^{3/8}\left(\frac{\ell}{Dt}\right)^{5/8}\,, \quad N\sim n_0^{3/4}\left(\frac{\ell}{Dt}\right)^{1/4}
\end{equation}
with logarithmic factor
\begin{equation}
\ell = \ln\!\left(\frac{Dt a^8}{n_0^3}\right)
\end{equation}

Generally for $d<2$ the proper generalization of \eqref{Nc:1d} reads (see \cite{book} for such arguments)
\begin{equation}
\label{Ncd}
\frac{dN}{dt} \sim - \frac{c_1}{T} \sim -Dc_1^{1+2/d}
\end{equation}
The same arguments as before give $c_m\propto m^{d/(d+2)}$ and after the same steps as above one gets the announced asymptotic behaviors \eqref{c1:below} and \eqref{N:below}.

\section{Final States and Evolution in Finite Systems}
\label{sec:FS}

Here we explore the ultimate fate of {\em finite} systems undergoing a maximally assortative exchange process. In this setting, the difference between maximally assortative and ordinary exchange processes is even more pronounced than for infinite systems. Indeed, in ordinary exchange processes all mass eventually accumulates in a single cluster. In a maximally assortative exchange process in a finite system, the final outcome is a jammed state containing clusters of different masses, so the exchange is no longer possible. 

\subsection{Final states}

A state $(m_1,\ldots,m_p)$ with cluster masses satisfying
\begin{equation}
\label{jammed}
1\leq m_1<\ldots<m_p, \qquad m_1+\ldots+m_p = \mathcal{M}
\end{equation}
is a jammed state of the system with total mass $\mathcal{M}$. The number of jammed states $J_\mathcal{M}$ increases with $\mathcal{M}$.  For small $\mathcal{M}$ one easily computes these numbers by hand; Table~\ref{Tab:JN} shows these numbers in the range $\mathcal{M}\leq 20$.

\begin{table}
\begin{tabular}{| c | c | c | c | c | c | c | c | c | c | c | c | c | c | c | c | c | c | c | c | c |}
\hline
$\mathcal{M}$     & 1 & 2 & 3 & 4 & 5 & 6 & 7 & 8 & 9 & 10 & 11 & 12 & 13 & 14 & 15 & 16 & 17 & 18 & 19 & 20\\ 
\hline
$J_\mathcal{M}$ & 1 & 1 & 2 & 2 & 3 & 4 & 5 & 6 & 8 & 10 & 12 & 15 & 18 & 22 & 27 & 32 & 38 & 46 & 54 & 64\\
\hline  
\end{tabular}
\caption{The number $J_\mathcal{M}$ of jammed states for  $1\leq \mathcal{M}\leq 20$. }
\label{Tab:JN}
\end{table} 

Contemplating about $J_\mathcal{M}$, i.e. the number of solutions of \eqref{jammed}, one realizes that $J_\mathcal{M}$ is the number of partitions of $\mathcal{M}$ into distinct parts. Such partitions appear in combinatorics \cite{Andrews}, often under the name of strict partitions; recently they have been also called Fermi partitions \cite{Vershik}. Strict partitions were first studied by Euler (see \cite{Euler:book}) who expressed the generating function for such partitions through an infinite product
\begin{equation}
\label{Euler:Fermi}
\sum_{\mathcal{M}\geq 0} J_\mathcal{M}\,Q^\mathcal{M} = \prod_{k\geq 1}(1+Q^k)
\end{equation}
(Here we have used the convention $J_0=1$.) Using \eqref{Euler:Fermi} and analyzing the $Q\to 1$ behavior one can extract the asymptotic behavior:  $\ln J_\mathcal{M}\simeq \pi\sqrt{\mathcal{M}/3}$ as $\mathcal{M} \to \infty$.  A more comprehensive analysis \cite{Andrews} gives the Ramanujan asymptotic formula
\begin{equation}
\label{Raman}
J_\mathcal{M} \simeq \frac{1}{4\cdot 3^{1/4}\,\mathcal{M}^{3/4}}\,\exp\!\left[\pi\sqrt{\frac{\mathcal{M}}{3}}\right]
\end{equation}

Despite of this growth of the total number of jammed states, the fate of the system is surprisingly deterministic, e.g., for the most natural initial condition when all clusters are initially monomers the final state is unique. This outcome is universal---the details of the exchange process are irrelevant, only the requirement that it is maximally assortative matters. Furthermore, the final state remains the same for many other initial conditions, e.g. if the initial number  $N_m(0)$ of clusters of mass $m$ satisfies $N_m(0)>0$ for all $m=1,\ldots,m_\text{max}$ and $N_m(0)=0$ for $m>m_\text{max}$; only if the initial mass distribution has big `holes' more complicated jammed states may arise. 

The final state is particularly simple when the initial mass is a triangular number, $\mathcal{M}=T_n=n(n+1)/2$ with arbitrary integer $n$. In this case, $(1,2,\ldots,n)$ is the final state. If the initial mass is not a triangular number, $T_{n-1} < \mathcal{M} < T_n$, the final state differs from $(1,2,\ldots,n)$ by a single hole: If we parametrize $\mathcal{M} = T_n - \ell$ with some $1\leq \ell<n$,  then the final state is 
\begin{equation}
\label{jammed-ell}
(1,\ldots,\ell-1,\widehat{\ell}, \ell+1,\ldots,n)
\end{equation}
where $\widehat{\ell}$ implies that the cluster with mass $\ell$ is absent. As an example, take $\mathcal{M} = 18 = T_6 - 3$. Equation \eqref{jammed-ell} tells us that the final state is $(1,2,4,5,6)$.

\subsection{Completion time}

The evolution towards the final state depends on the details of the dynamics, and even for the fixed dynamics the duration varies from realization to realization, that is, the time $t_\text{final}$ to reach the final state is a random variable. First, we estimate the completion time for the simplest maximally assortative exchange process with mass-independent migration rates. When $\mathcal{M}\gg 1$, the behavior is initially the same as the behavior of the infinite system (Sect.~\ref{sec:const}). Therefore the total number $N_m$ of clusters of mass $m$ is 
\begin{equation}
\label{Nmt:0}
N_m(t) = \frac{\mathcal{M}}{\sqrt{t}}\,G(y)
\end{equation}
with $y=m/(80t)^{1/4}$ and $G(y)$ given by \eqref{G:0}. These formulas formally apply when $N_m\gg 1$, but we can employ them up to $N_m = O(1)$ in estimates. Thus we use the criterion $\mathcal{M}\sim \sqrt{t_\text{final}}$ to estimate 
\begin{equation}
t_\text{final} \sim \mathcal{M}^2
\end{equation}

A similar argument for maximally assortative exchange processes with algebraic migration rates $K(m)=m^a$ gives $\mathcal{M}\sim t_\text{final}^{2/(4-a)}$. Therefore the completion time scales with total mass according to  
\begin{equation}
t_\text{final} \sim \mathcal{M}^{2-a/2}
\end{equation}
This is valid when $a\leq 1$. The population of monomers exceeds the population of clusters of any other mass when $a>1$, so we cannot use continuum predictions when $N_1 = O(1)$ since other cluster densities are negligible at such times and the continuum approach cannot be trusted in this domain. 

\subsection{The extremal model ($a=\infty$)}
\label{sec:FS-extreme}

Let us look at the maximally assortative exchange process with $a=\infty$, equivalently a process with infinitely fast migration rates $K(m)=\infty$ for all $m\geq 2$. In this extremal model the composition of the system is remarkably simple: When $t<t_\text{final}$, we still have a lot of monomers, $N_1\gg 1$, while the rest of the population is composed like \eqref{jammed-ell}, namely $N_m=1$ for $2\leq m\leq m_0(t)$ with at most a single hole inside. 

To describe the evolution we notice that the merging of two monomers takes a positive time and it may trigger an avalanche of other exchanges which proceed instantaneously. Symbolically $1\oplus 1\to 2$ and there will be no other instantaneous exchanges if in the preceding configuration the dimer was absent; otherwise $2\oplus 2\to (1,3)$ will occur, perhaps followed by a longer avalanche of instantaneous exchanges. Focusing on the population of monomers we have $N_1\to N_1-2$ in the first case and $N_1\to N_1-1$ in the second. In the long time limit a hole (if it exists) is usually far away, so that $N_1\to N_1-1$ dominates. As long as $N_1\gg 1$, we can use the rate equation 
\begin{equation}
\label{c1:inf}
\frac{dc_1}{dt} = - c_1^2
\end{equation}
for the monomer density $c_1=N_1/\mathcal{M}$. This is very simple, but conceptually remarkable result. Recall that \eqref{exchange:a} predicts that the density of monomers satisfies 
\begin{equation}
\label{c12:inf}
\frac{dc_1}{dt} = 2^a c_2^2 - 2c_1^2
\end{equation}
It is not immediately obvious how to interpret the first term on the right-hand side of \eqref{c12:inf} when $a=\infty$. The above analysis shows that we must drop this term and divide by two the pre-factor of the second term. This reminds taking the zero-viscosity limit in turbulence, e.g. in Burgers turbulence one keeps the dissipation rate finite and justifies it by appearance of shocks (see e.g. \cite{BD99,E00,Khanin07,Anna14}). 

Solving \eqref{c1:inf} we get $c_1=(1+t)^{-1}$ and hence
\begin{subequations}
\begin{equation}
\label{N1:inf}
N_1 = \frac{\mathcal{M}}{1+t}
\end{equation}
The rest of the mass distribution is
\begin{equation}
\label{Nm:inf}
N_m = 1, \quad 2\leq m\leq m_0(t)\simeq \sqrt{2\mathcal{M}\,\frac{t}{1+t}}
\end{equation}
\end{subequations}
The largest mass $m_0(t)$ is established from the requirement of mass conservation. In the interesting $1\ll t \ll \mathcal{M}$ time range where we can employ deterministic rate equations the fraction of mass carried by monomers decreases as $t^{-1}$.  We also notice that the total number of clusters 
\begin{equation}
\label{N:finite}
\mathcal{N} = N_1 + m_0(t) = \frac{\mathcal{M}}{1+t} + \sqrt{2\mathcal{M}\,\frac{t}{1+t}}
\end{equation}
has an interesting behavior: The monomers provide the dominant contribution when $t \ll \sqrt{\mathcal{M}}$, while for $t\gg \sqrt{\mathcal{M}}$ the total number of clusters saturates to 
\begin{equation}
\label{N:inf}
\mathcal{N}_\text{final} = \sqrt{2\mathcal{M}}
\end{equation}

Combining \eqref{N1:inf} and the criterion $N_1 = O(1)$ we estimate the completion time
\begin{equation}
t_\text{final} \sim \mathcal{M}
\end{equation}
It is worth mentioning that for the extremal maximally assortative exchange process with $a=\infty$ one can derive much more precise results about the completion time. Indeed, in the most interesting case when $\mathcal{M}\gg 1$, we established that the dominant channel describing the decrease of monomers is $N_1\to N_1-1$, i.e. monomers effectively undergo the coalescence process: $A_1+A_1\to A_1$. This stochastic process is well-understood and the probability distribution for the completion time is known (see \cite{book}). For instance, the leading behaviors of the two basic moments of $t_\text{final}$ are
\begin{equation}
\langle t_\text{final}\rangle = \mathcal{M}, \qquad \frac{\langle t_\text{final}^2\rangle}{\langle t_\text{final}\rangle^2} = \frac{\pi^2}{3}-2
\end{equation}
Thus fluctuations do not die even in the thermodynamic limit $\mathcal{M}\to \infty$. To appreciate it suffices to note that the process $N_1\to N_1-1$ occurs with rate $N_1(N_1-1)/\mathcal{M}$, so its average duration is $\frac{\mathcal{M}}{N_1(N_1-1)}$. Therefore last steps when $N_1=O(1)$ take time $O(\mathcal{M})$ and this explains the non-self-averaging nature. Up until the very end, however, the evolution is essentially deterministically. 

Overall, the extremal maximally assortative exchange process exhibits a very peculiar behavior. There is no gel (which by definition is a giant cluster containing a finite fraction of mass of the entire system). On the other side, in non-gelling systems or non-gelling phases, the largest cluster usually has a mass of the order of $\ln \mathcal{M}$, while in  the extremal maximally assortative exchange process there are numerous clusters with masses of the order of $\sqrt{\mathcal{M}}$, and these clusters actually contain most of the mass is actually.

\section{Exchange Processes Driven by a Localized Input of Monomers}
\label{sec:input}

Reaction-diffusion processes driven by localized input often occur in Nature and they are also used in various industrial applications. Some of these processes involve a few species of atoms; as examples we mention electropolishing \cite{EP}, dissolution \cite{D87}, corrosion \cite{KM91}, and erosion \cite{SBG}. These processes are rather tractable  \cite{Lar,PLK:Stefan,IDLA_1,IDLA_2} and well understood. Other processes involve numerous interacting sub-species, e.g. clusters in aggregation \cite{Sid89,PLK:3particle,PLK:source,Asymmetric:source,Kirone:source,PLK12} and ordinary mass exchange \cite{PK_exchange}; the analysis of these systems is much more challenging and usually relies on non-rigorous tools. 

Here we study maximally assortative exchange processes driven by a localized input. The densities $c_m(\mathbf{r},t)$ obey an infinite system of non-linear coupled PDEs
\begin{eqnarray}
\label{loc:input}
\frac{\partial c_m}{\partial t} &=& 
K(m+1) c_{m+1}^2-2K(m) c_m^2+K(m-1) c_{m-1}^2 \nonumber\\
&+&D_m\nabla^2 c_m  + J \delta_{m,1}\delta(\mathbf{r})\theta(t)
\end{eqnarray}
The terms on the top line on the right-hand side of \eqref{loc:input} account for exchange. The first term on the bottom line describes mixing due to diffusion and the following term represents the input of monomers source  ($J$ is the strength of the monomer flux) at the origin. We are interested at the behavior on distances greatly exceeding the size of the region where monomers are injected and hence we model the flux using the delta function $\delta(\mathbf{r})$. The source is turned at $t=0$, as specified by $\theta(t)$ on the right-hand side of  \eqref{loc:input}; before that moment the system is assumed to be empty. In the following $t>0$ and we do not explicitly write $\theta(t)=1$. 

\subsection{Mass-independent hopping rates}

Here we study the model with mass-independent migration rates and diffusion coefficients. For the diffusion-controlled point cluster exchange process on the lattice, the migration rates are proportional to the corresponding hopping rates, $K(m)\sim D_m$, so if diffusion coefficients are mass-independent the  migration rates are also mass-independent. Equations \eqref{loc:input} for this model become
\begin{equation}
\label{loc:input-const}
\frac{\partial c_m}{\partial t} = \nabla^2 c_m + c_{m+1}^2-2c_m^2+c_{m-1}^2 + J \delta_{m,1}\delta(\mathbf{r})
\end{equation}
where we have set $K(m)=1$ and $D_m=1$. 

The mass density $M(\mathbf{r},t)=\sum_{m\geq 1} m c_m(\mathbf{r},t)$ is now spatially dependent and it also depends on time. The mass is not affected by the exchange, so it satisfies the diffusion equation with a localized source
\begin{equation}
\label{loc:mass-const}
\frac{\partial M}{\partial t} = \nabla^2 M  + J\delta(\mathbf{r})
\end{equation}
which can be solved in arbitrary dimension.  

\subsubsection{Three dimensions}
\label{sec:3d-long}

In the most physically relevant three-dimensional case the rate equation approach is applicable. An extra simplification is that in three dimensions (and generally when $d>2$), the mass density is stationary; more precisely, the mass density coincides with Coulomb potential generated by `charge' $J$, viz.
\begin{equation}
\label{Mr}
M = \frac{J}{4\pi r}
\end{equation}
Since the source is turned on at $t=0$ and clusters propagate diffusively, the stationarity ceases to hold when $r\sim \sqrt{t}$, and $M(r,t)$ quickly approaches to zero as $r t^{-1/2}$ increases. 

Other natural quantities do not even satisfy closed equations. For instance, the total cluster density $N(\mathbf{r},t)=\sum_{m\geq 1} c_m(\mathbf{r},t)$ evolves according to 
\begin{equation}
\label{loc:density-const}
\frac{\partial N}{\partial t} = \nabla^2 N  - c_1^2 + J\delta(\mathbf{r})
\end{equation}

It is reasonable to assume that in the long time limit the densities become stationary. More precisely, they are stationary as long as the distance is not too far from the source, namely $r\ll \sqrt{t}$.  In the stationary regime in three dimensions \eqref{loc:density-const} becomes 
\begin{equation}
\label{Nr3}
\frac{1}{r^2}\,\frac{d}{dr}\left(r^2\,\frac{dN}{dr}\right) - c_1^2 + J\delta(\mathbf{r})=0
\end{equation}
Further, in the stationary regime in three dimensions  Eqs.~\eqref{loc:input-const}  read
\begin{equation}
\label{cmr3}
\frac{1}{r^2}\,\frac{\partial }{\partial r}\left(r^2\,\frac{\partial  c}{\partial r}\right) + \frac{\partial^2 }{\partial m^2}\,c^2 = 0
\end{equation}
where we have replaced $c_{m+1}^2-2c_m^2+c_{m-1}^2$ by the second derivative which should be asymptotically exact when $m\gg 1$. We seek a solution to \eqref{cmr3} in a scaling form
\begin{equation}
\label{scaling:r}
c(m,r) = c_m(r) = r^{-2\beta-1} \Phi(x), \qquad x=\frac{m}{r^\beta}
\end{equation}
The pre-factor $r^{-2\beta-1}$ is chosen to be consistent with \eqref{Mr}. Indeed 
\begin{equation}
M(r) = \sum_{m\geq 1} m c_m(r)\simeq r^{-1}\int_0^\infty dx\,x\Phi(x)
\end{equation}
has correct spatial dependence, and the complete match is obtained if 
\begin{equation}
\label{match}
\int_0^\infty dx\,x\Phi(x)=\frac{J}{4\pi}
\end{equation}

By inserting \eqref{scaling:r} into \eqref{cmr3} we deduce $\beta=1/4$ and
\begin{equation}
\label{Phi:1}
16 (\Phi^2)'' + x^2 \Phi''+9x\Phi'+12\Phi = 0
\end{equation}
This non-linear second-order ODE with non-constant coefficients is soluble. First we notice that \eqref{Phi:1} admits an integrating factor: Multiplying \eqref{Phi:1} by $x$ we obtain
\begin{equation*}
[16x(\Phi^2)' - 16\Phi^2 + x^3 \Phi'+6x^2\Phi]' = 0
\end{equation*}
which we integrate and write the outcome as
\begin{equation}
\label{Phi:2}
16\,\frac{d}{dx}\!\left(\frac{\Phi^2}{x}\right) + x\,\frac{d \Phi}{dx}+6\Phi = 0
\end{equation}
(The integration constant was to zero to assure that $\Phi(x)$ vanishes as $x\to\infty$.) We simplify the first-order ODE \eqref{Phi:2} by making the transformation
\begin{equation}
\label{Phi-Psi}
\Phi(x) = \sqrt{x}\,\Psi(x)
\end{equation}
Using $Y=x^{3/2}$ instead of $x$ we find that $\psi(Y)=\Psi(x)$ satisfies 
\begin{equation}
\label{Psi:eq}
3\frac{d\psi}{dY} + \frac{13\psi}{32\Psi + Y} =0
\end{equation}
This equation simplifies if instead of $\psi(Y)$ we consider the inverse function $Y(\psi)$: 
\begin{equation*}
\frac{13}{3}\,\frac{dY}{d\psi} = -32 - \frac{Y}{\psi}
\end{equation*}
Solving this equation we arrive at
\begin{equation}
\label{implicit:Y}
Y = 6\big(C\,\psi^{-3/13} - \psi\big)
\end{equation}
where $C$ is an integration constant. Returning to the original variables we obtain an implicit solution
\begin{equation}
\label{implicit}
\frac{x^2}{6} = C\left[\frac{x^8}{\Phi^3}\right]^{\frac{1}{13}} - \Phi 
\end{equation}

\begin{figure}
\centering
\includegraphics[width=7.77cm]{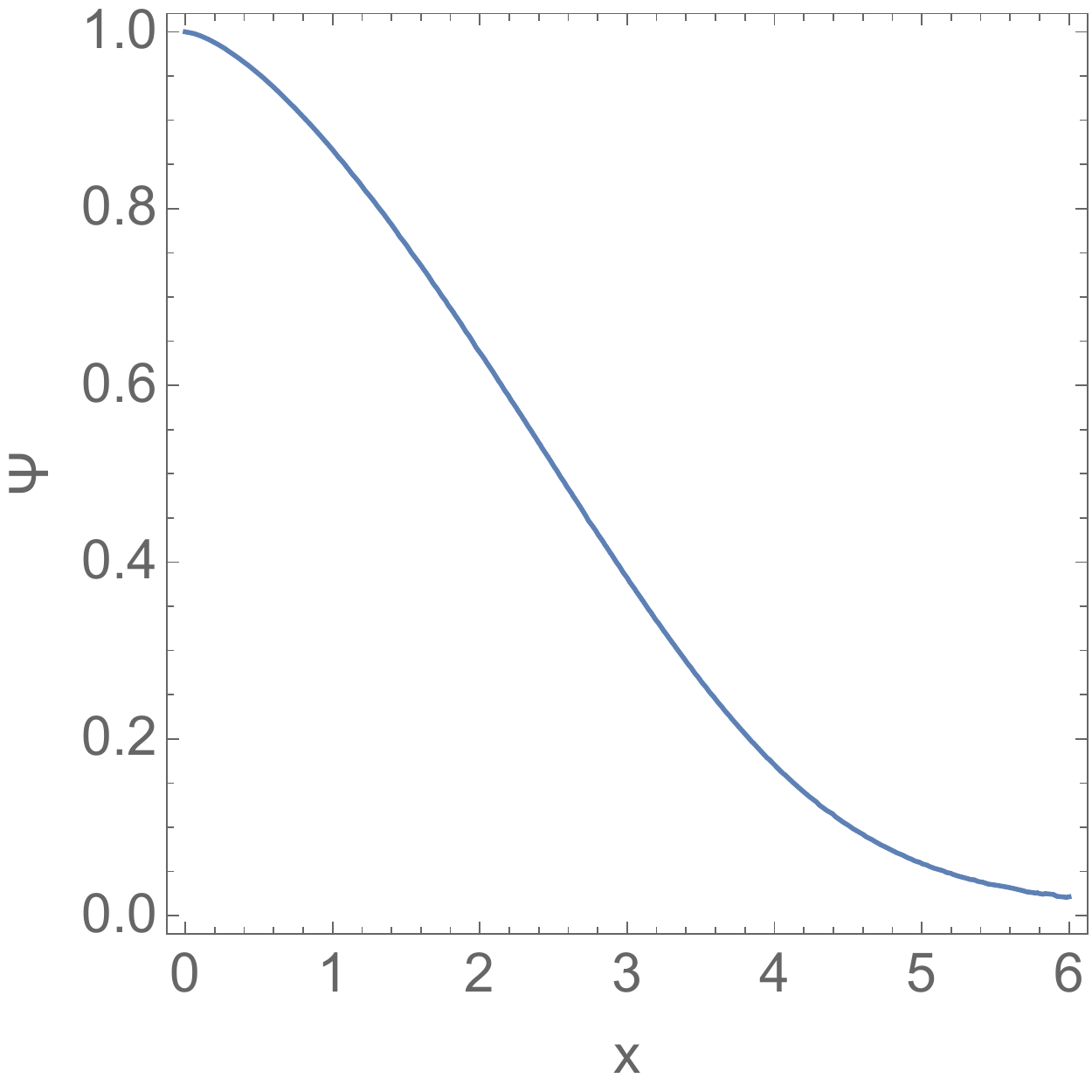}
\caption{The renormalized scaled mass distribution $\Psi(x)$ defined via \eqref{Phi-Psi}. The parameter $C$ in \eqref{implicit} is chosen to be $C=1$; this corresponds to the flux strength $J=A^{-1}$ with $A$ appearing in \eqref{C-A}.}
\label{Fig:Psi}
\end{figure}

The limiting behaviors of the scaled mass distribution are (see also Fig.~\ref{Fig:Psi})
\begin{equation}
\label{Phi:limits}
\Phi \simeq
\begin{cases}
C^{\frac{13}{16}}\,\sqrt{x}       & x\to 0\\
(6C)^{\frac{13}{3}}\,x^{-6}      & x\to \infty
\end{cases}
\end{equation}

The constant $C$ is fixed by \eqref{match}. To compute the integral in \eqref{match} we first re-write it as
\begin{equation}
\label{int}
\int_0^\infty dx\,x\Phi(x) = \frac{2}{3} \int_0^\infty dY\,Y^\frac{2}{3}\psi(Y)
\end{equation}
Equation \eqref{implicit} shows that $0\leq \psi\leq \psi_0=C^{13/16}$. We can now compute the integral in \eqref{int} using integration by parts, Eq.~\eqref{implicit} and straightforward transformations: 
\begin{eqnarray}
\label{long-integral}
 \frac{2}{3}\int_0^\infty dY\,Y^\frac{2}{3}\psi(Y) & = &  \frac{2}{5}\int_0^{\psi_0} d\psi\, Y^\frac{5}{3} \nonumber \\
         & = & \frac{2}{5}\,6^\frac{5}{3}\int_0^{\psi_0} d\psi\, \big(C\,\psi^{-3/13} - \psi\big)^\frac{5}{3} \nonumber \\
         & = & \frac{2}{5}\,6^\frac{5}{3}\,\psi_0^\frac{8}{3}\int_0^1 du\, \big(u^{-3/13} - u\big)^\frac{5}{3}  \nonumber \\
         & = & 6^\frac{5}{3}\,C^{\frac{13}{6}}\, \frac{13\sqrt{\pi}\,\Gamma\!\left(\frac{8}{3}\right)}{40\,\Gamma\!\left(\frac{19}{6}\right)}
\end{eqnarray}
Combining this with \eqref{match} we express the amplitude $C$ through the flux:
\begin{equation}
\label{C-A}
C=(AJ)^{6/13}, \quad A = \frac{10\,\Gamma\!\left(\frac{19}{6}\right)}{13\pi^{3/2}\,6^{5/3}\,\Gamma\!\left(\frac{8}{3}\right)}
\end{equation}

Let us also compute the cluster density. We have
\begin{equation}
N(r) = \sum_{m\geq 1} c_m(r)\simeq r^{-5/4}\int_0^\infty dx\,\Phi(x)
\end{equation}
The last integral is computed using the same tricks as in the computation in Eq.~\eqref{long-integral}. We get
\begin{eqnarray*}
\int_0^\infty dx\,\Phi(x) &= &  \frac{2}{3} \int_0^\infty dY\,\psi(Y) \\
                                    & = &  \frac{2}{3}\int_0^{\psi_0} d\psi\, Y \\
                                    & = &  4\int_0^{\psi_0} d\psi\, \big(C\,\psi^{-3/13} - \psi\big) \\
                                    & = &  \frac{16}{5}\,\psi_0^2 = \frac{16}{5}\,C^\frac{13}{8}
\end{eqnarray*}
leading to
\begin{subequations}
\begin{equation}
\label{Nr:3d}
N(r) =  \tfrac{16}{5}\,(AJ)^{3/4}\, r^{-5/4}
\end{equation}
The monomer density is found from \eqref{scaling:r} and the $x\to 0$ asymptotic of $\Phi(x)$, see \eqref{Phi:limits}, to give
\begin{equation}
\label{c1r:3d}
c_1(r) =  (AJ)^{3/8}\, r^{-13/8}
\end{equation}
\end{subequations}
As a useful consistency check we note that \eqref{Nr:3d} and \eqref{c1r:3d} agree with \eqref{Nr3}. 

\subsubsection{High dimensions}

The three-dimensional case is most relevant, but it is amusing to explore the behavior in dimension $d=4$ and higher. It turns out that the exchange is barely relevant at the `upper' critical dimension $d=d^c=4$ and asymptotically irrelevant in higher dimensions. To see this let us treat $d$ as a continuous parameter. The rate equation approach is generally applicable when $d>2$. The mass density is stationary and given by  
\begin{equation}
\label{Mr:d}
M = \frac{J}{(d-2)\Omega_d\, r^{d-2}}
\end{equation}
where $\Omega_d = \frac {2\pi^{d/2}}{\Gamma(d/2)}$ is the `area' of unit sphere $\mathbb{S}^{d-1}$. This expression suggests that the relevant generalization of the three-dimensional scaling form \eqref{scaling:r} is
\begin{equation}
\label{scaling:rd}
c(m,r) = c_m(r) = r^{-2\beta-d+2} \Phi(x), \qquad x=\frac{m}{r^\beta}
\end{equation}
Plugging this form into 
\begin{equation}
\label{cmrd}
\frac{1}{r^{d-1}}\,\frac{\partial }{\partial r}\left(r^{d-1}\,\frac{\partial  c}{\partial r}\right) + \frac{\partial^2 }{\partial m^2}\,c^2 = 0
\end{equation}
we deduce $\beta=1-d/4$ and determine the scaled mass density (see Appendix \ref{sec:ap-high}). 

The small mass behavior is again $\Phi\sim \sqrt{x}$, and the monomer density decays according to
\begin{equation}
\label{c1r:d}
c_1(r) =  (A_d J)^{3/8}\, r^{-(3d+4)/8}
\end{equation}
The cluster density is given by  
\begin{equation}
\label{Nr:d}
N(r) =  \tfrac{16}{(3d-4)(4-d)}\,(A_d J)^{3/4}\, r^{-(3d-4)/4}
\end{equation}
The decay law \eqref{Nr:d} can be extracted from \eqref{c1r:d} and 
\begin{equation*}
\frac{1}{r^{d-1}}\,\frac{d}{dr}\left(r^{d-1}\,\frac{dN}{dr}\right) - c_1^2 = 0
\end{equation*} 

The scaling form \eqref{scaling:rd} is applicable when $2<d<4$. Indeed, when $d\leq d_c=2$ we cannot use mean-field rate equations. The upper bound $d<d^c=4$ is obvious from the above formulas, e.g. the exponent $\beta=1-d/4$ must be positive, yet it vanishes at $d=4$ and becomes negative when $d>4$. In sufficiently high dimensions, $d>4$, clusters essentially do not `see' each other. More precisely, some exchange processes occur near the source, but then clusters hardly meet. Therefore both the monomer density and the total cluster density decay in the similarly to the mass density:
\begin{equation}
\label{c1Nr:blind}
c_1 \sim r^{-(d-2)}, \qquad N \sim r^{-(d-2)}
\end{equation}

The exponent $d-2$ approaches to two as $d\to 4$. Since $\beta=0$  at the upper critical dimension $d=d^c=4$, we 
anticipate that $m$ scales logarithmically. Thus we seek the mass distribution in the form 
\begin{equation}
c_m(r) = r^{-2}C_m(\rho), \quad \rho=\ln r
\end{equation}
Plugging this ansatz into the governing equations 
\begin{equation}
\frac{1}{r^3}\,\frac{d}{dr}\left(r^3\,\frac{dc_m}{dr}\right) +c_{m-1}^2 - 2c_m^2 + c_{m+1}^2 = 0
\end{equation}
we obtain
\begin{equation}
\label{Cmr:long}
2\frac{d C_m}{d \rho} + \frac{d^2 C_m}{d \rho^2}  = C_{m-1}^2 - 2C_m^2 + C_{m+1}^2 
\end{equation}
The interesting behavior occurs far from the source where the second term on the right-hand size of \eqref{Cmr:long} is negligible in comparison with the first term. (This is asymptotically true; however, the ratio of these two terms vanishes as $\rho^{-1}$, and since $\rho=\ln r$ the ratio decays very slowly.) Dropping the second term on the right-hand size of \eqref{Cmr:long} we arrive at
\begin{equation}
\label{Cmr:short}
2\frac{d C_m}{d \rho}   = C_{m-1}^2 - 2C_m^2 + C_{m+1}^2 
\end{equation}
This set of equations can be identified with \eqref{exchange:const} after the transformation
\begin{equation}
C_m(\rho)=\frac{J}{4\pi^2}\,c_m\!\left(\frac{J\rho}{8\pi^2}\right)
\end{equation}
which also matches \eqref{mass} with $\sum_{m\geq 1}mC_m = \frac{J}{4\pi^2}$ following from \eqref{Mr:d} at $d=4$. Using previous results we deduce that when $y<1$ the mass density distribution is given by 
\begin{equation}
c_m(r) = \sqrt{\frac{5J}{18\pi^2}}\,\frac{\sqrt{y}-y^2}{r^2 \rho^{1/2}}, \quad 
y = m\left(\frac{10J\rho }{\pi^2}\right)^{-1/4}
\end{equation}
In particular
\begin{subequations}
\begin{align}
\label{Nr:4}
N(r) & = B\,\frac{J^{3/4}}{r^2 \rho^{1/4}}\\
\label{c1r:4}
c_1(r)& = B_1\,\frac{J^{3/8}}{r^2 \rho^{5/8}}
\end{align}
\end{subequations}
with
\begin{equation*}
B_1= \frac{5^{3/8}}{3\cdot 2^{5/8}\cdot \pi^{3/4}}, \qquad B = \frac{5^{3/4}}{9\cdot 2^{1/4}\cdot \pi^{3/2}}
\end{equation*}

\subsubsection{Low dimensions}

When $d\leq 2$ the rate equation approach becomes erroneous. There are no closed form exact equations for cluster densities, but modified rate equations provide qualitatively correct results and lead to exact scaling. We now outline the results for $d=1$ and $d=2$. 

In one dimension, we seek the scaling solution in the form $c(m,r)=m^{-\alpha}F(m/r^\beta)$. Plugging this ansatz into the analog of \eqref{cmrd}, namely $\frac{\partial^2 c}{\partial r^2}+\frac{\partial^2 c^3}{\partial m^2}=0$, we deduce the relation $\beta=(1+\alpha)^{-1}$ between the scaling exponents. Estimating $\sum_{m\geq 1}mc_m\sim r^{(2-\alpha)\beta}$ and noting that it should scale as $r$ we deduce the second relation $\beta(2-\alpha)=1$. From these relations $\alpha=\frac{1}{2}$ and $\beta=\frac{2}{3}$. One can also establish proper powers of the source strength (omitted above). The scaling form reads  
\begin{equation}
\label{scaling:r1}
c_m(r) =  \sqrt{\frac{J}{m}}\,\Phi(x), \qquad x=\frac{m}{J^{1/3}r^{2/3}}
\end{equation}
Using this expression we estimate the cluster density 
\begin{subequations}
\begin{equation}
\label{Nr:1d}
N(r)\sim J^{2/3}r^{1/3}
\end{equation}
Equations \eqref{scaling:r1} and \eqref{Nr:1d} are consistent  with $\frac{d^2 N}{dr^2}\sim c_1^3$ if $\Phi(x)\sim x^{5/6}$ as $x\to 0$. Thus
\begin{equation}
\label{c1r:1d}
c_1(r) \sim J^{2/9}r^{-5/9}
\end{equation}
\end{subequations}

In two dimensions, we obtain
\begin{subequations}
\begin{align}
\label{Nr:2d}
N(r)&\sim J^{3/4} \rho^{1/4} r^{-1/2}\\
\label{c1r:2d}
c_1(r) &\sim J^{3/8} \rho^{5/8} r^{-5/4}
\end{align}
\end{subequations}
where we again shortly write $\rho=\ln r$. 

\subsubsection{Total numbers of monomers and clusters}

The total number of monomers $\mathcal{C}_1(t)$ is estimated by integrating the stationary density till $r=\sqrt{t}$. Thus
\begin{equation*}
\mathcal{C}_1(t) \sim \int_0^{\sqrt{t}} dr\,r^{d-1} c_1(r)
\end{equation*}
Using \eqref{c1r:1d}, \eqref{c1r:2d}, \eqref{c1r:3d} and \eqref{c1r:4} we obtain
\begin{equation}
\label{monomers:input}
\mathcal{C}_1 \sim
\begin{cases}
J^{2/9} t^{2/9}                       & d=1\\
J^{3/8} t^{3/8}  (\ln t)^{5/8}    & d=2\\
J^{3/8} t^{11/16}                   & d=3\\
J^{3/8} t  (\ln t)^{-5/8}           & d=4\\
J t                                         & d>4
\end{cases}
\end{equation}
Similarly the total number of clusters is estimated from 
\begin{equation*}
\mathcal{N}(t) \sim \int_0^{\sqrt{t}} dr\,r^{d-1} N(r)
\end{equation*}
Using \eqref{Nr:1d}, \eqref{Nr:2d}, \eqref{Nr:3d} and \eqref{Nr:4} we obtain
\begin{equation}
\label{clusters:input}
\mathcal{N} \sim
\begin{cases}
J^{2/3} t^{2/3}                       & d=1\\
J^{3/4} t^{3/4}  (\ln t)^{1/4}    & d=2\\
J^{3/4} t^{7/8}                       & d=3\\
J^{3/4} t  (\ln t)^{-1/4}            & d=4\\
J t                                         & d>4
\end{cases}
\end{equation}

\subsection{Mass-dependent rates}

Diffusion coefficients generally decrease with mass. An algebraic decay, $D_m\sim m^{-\nu}$, often occurs, e.g., the mobility exponents $\nu=1$ and $\nu=3/2$ arise in problems involving two-dimensional clusters \cite{einstein}.  For the diffusion-controlled point cluster exchange processes on the lattice $K(m)\sim D_m$ suggesting to study the models with $D_m\sim K(m)\sim m^{-\nu}$. The behavior of such models driven by a local source can be treated using the same scheme as before, namely assuming the emergence of a stationary mass distribution and the validity of scaling. 

As a concrete example, let us consider the model with $D_m= K(m)=m^{-1}$. The rate equations read
\begin{eqnarray}
\label{loc:input-inverse}
\frac{\partial c_m}{\partial t} &=&  (m+1)^{-1}c_{m+1}^2-2m^{-1}c_m^2+ (m-1)^{-1}c_{m-1}^2\nonumber\\
& + & m^{-1}\nabla^2 c_m + J \delta_{m,1}\delta(\mathbf{r})
\end{eqnarray}
The mass density now varies according to 
\begin{equation}
\label{loc:mass-inverse}
\frac{\partial M}{\partial t} = \nabla^2 N  + J\delta(\mathbf{r})
\end{equation}
In the most physically relevant three-dimensional case, Eq.~\eqref{loc:mass-inverse} gives a simple expression 
\begin{equation}
\label{Nr:inv}
N = \frac{J}{4\pi r}
\end{equation}
for the cluster density in the long time limit. 

Let us explore the stationary regime in three dimensions in more detail. 
We simplify Eqs.~\eqref{loc:input-inverse} to
\begin{equation}
\label{cmr3:inv}
\frac{1}{r^2}\,\frac{\partial }{\partial r}\left(r^2\,\frac{\partial  c}{\partial r}\right) + m\,\frac{\partial^2 }{\partial m^2}\,\frac{c^2}{m} = 0
\end{equation}
and seek a solution to \eqref{cmr3:inv} in a scaling form
\begin{equation}
\label{scaling:inv}
c(m,r) = c_m(r) = r^{-\beta-1} \Phi(x), \qquad x=\frac{m}{r^\beta}
\end{equation}
The pre-factor $r^{-\beta-1}$ is consistent with \eqref{Nr:inv}. Indeed
\begin{equation}
N(r) = \sum_{m\geq 1} c_m(r)\simeq r^{-1}\int_0^\infty dx\,\Phi(x)
\end{equation}
assures the correct spatial decay of the cluster density, and the constraint 
\begin{equation}
\label{match:inv}
\int_0^\infty dx\,\Phi(x)=\frac{J}{4\pi}
\end{equation}
provides the complete match with \eqref{Nr:inv}. By inserting \eqref{scaling:inv} into \eqref{cmr3:inv} we deduce $\beta=1/3$ and
\begin{equation}
\label{Phi:inv}
9x (x^{-1}\Phi^2)'' + x^2 \Phi''+6x\Phi'+4\Phi = 0
\end{equation}
which is integrated to yield $9 (x^{-2}\Phi^2)' + \Phi'+4x^{-1}\Phi = 0$. The implicit solution to this equation reads
\begin{equation}
\label{implicit:inv}
\tfrac{1}{3}x^2 = C (x^6/\Phi)^{1/5}-\Phi
\end{equation}
with $C$ being an integration constant. The limiting behaviors of the scaled mass distribution are 
\begin{equation}
\Phi \simeq
\begin{cases}
C^{\frac{5}{6}}\,x            & x\to 0\\
(3C)^{5}\,x^{-4}              & x\to \infty
\end{cases}
\end{equation}
Using \eqref{implicit:inv} we compute the integral in \eqref{match:inv} and extract the amplitude $C=\left(\frac{J}{16\pi}\right)^{2/5}$. In particular,
\begin{equation}
c_1(r) = \left(\frac{J}{16\pi}\right)^{2/5} r^{-5/3}
\end{equation}

\begin{figure}
\centering
\includegraphics[width=7.77cm]{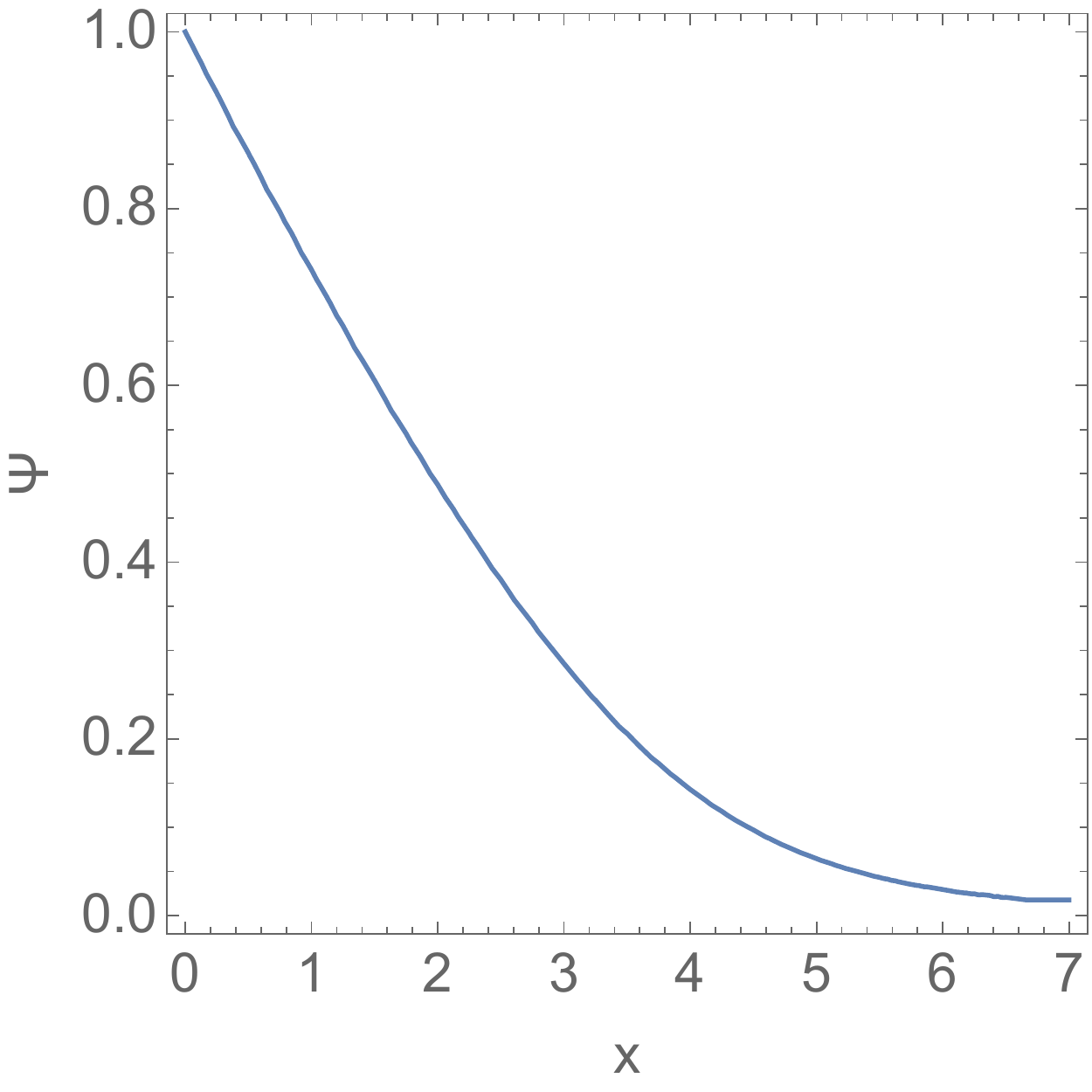}
\caption{The renormalized scaled mass distribution defined via $\Psi(x)=x^{-1}\Phi(x)$. The parameter $C$ in \eqref{implicit:inv} is chosen to be $C=1$; this corresponds to the flux strength $J=16\pi$.}
\label{Fig:Psi2}
\end{figure}

To determine the spatial size $R$ of the region where the densities have become stationary, we first compute the mass density:
\begin{equation*}
M(r) = \sum_{m\geq 1} mc_m(r)\simeq r^{-2/3}\int_0^\infty dx\,x\Phi(x)
\end{equation*}
The integral is calculated using \eqref{implicit:inv} to yield 
\begin{equation}
M(r) =   \frac{729}{56} \left(\frac{J}{16\pi}\right)^{4/3} r^{-2/3}
\end{equation}
Ignoring numerical factors we estimate the total mass 
\begin{equation*}
\mathcal{M} \sim \int_0^R dr\,r^2 M(r) \sim J^\frac{4}{3}\, R^\frac{7}{3}
\end{equation*}
Since $\mathcal{M} = Jt$ we have $R\sim J^{-1/7} t^{3/7}$. Using this result and \eqref{Nr:inv} we estimate the total number of clusters \begin{equation}
\mathcal{N}(t) \sim \int_0^R dr\,r^2 N(r) \sim J R^2 \sim J^\frac{5}{7}\, t^\frac{6}{7}
\end{equation}
Note also the asymptotic growth law for the total number of monomers
\begin{equation}
\mathcal{C}_1(t) \sim J^\frac{22}{105}\, t^\frac{4}{7}
\end{equation}

\section{Discussion}
\label{sec:disc}

Maximally assortative exchange processes are mathematically challenging and not a single one has been solved so far. We have shown that for a class of models with algebraic migration rates, $K(m)=m^a$, we relied on scaling to establish the most interesting asymptotic behaviors in the $a\leq 3$ range. It would be interesting to understand the behavior when $a>3$ where scaling is violate. 

When scaling holds, a single typical mass characterizes the mass distribution. The mass distribution in the extremal model ($a=\infty$) has two scales: $m=1$ corresponding to the monomers and the scale $m_0$, see \eqref{Nm:inf}, characterizing the rest of the system. This suggests that when $3<a<\infty$ there may be two scales, an inner region $m\sim t^{\beta_-}$ and much larger outer region with $t^{\beta_-}\ll m \sim t^{\beta_+}$. Mass distributions with two scales, and even three, scales have appeared in a few models of aggregation with uniform input, see \cite{BK91,Jose98,Jose99}. In the present situation, however, we haven't succeeded in establishing a consistent a boundary layer structure of the mass distribution. 

A strange feature of the mass distribution in the extremal model is that the outer scale $m_0$ is asymptotically independent on time, but depends on the total mass of the system: $m_0\simeq \sqrt{2\mathcal{M}}$. Thus for infinite systems, $\mathcal{M}=\infty$, the extremal model provides little insight for guessing the behavior when $3<a<\infty$, or perhaps the message is well hidden. Overall, the extremal model resembles taking the zero-viscosity limit in turbulence---the terms containing $a=\infty$ formally disappear, yet they affect the evolution.   

The behavior of maximally assortative exchange processes substantially differs from the behavior of ordinary exchange processes. To study the interpolation between these two extremes one can introduce parameter $r\in [0,1]$ measuring the degree of assortatitivity by postulating that the reaction channel \eqref{Aij} to operate only when $r\leq \frac{i}{j}\leq r^{-1}$. With this definition, $r=1$ corresponds to maximally assortative exchange processes and $r=0$ corresponds to ordinary exchange processes. The extreme behaviors are known for simple rates rates such as $K_{i,j}=(ij)^{a/2}$; ordinary exchange processes with these rates were studied in \cite{EP03}, while for maximally assortative exchange processes we recover the rates $K(m)=m^a$.  We know that e.g. the cluster density decays as
\begin{equation}
\label{Nr}
N\sim
\begin{cases}
t^{-1/(3-a)}    & \text{when} ~~r=0\\
t^{-1/(4-a)}     & \text{when} ~~r=1
\end{cases}
\end{equation} 
These asymptotic results are valid when $a<3$. One would like to understand how $r$ affects the decay law for the cluster density and behaviors of other quantities.

\appendix
\section{Mass density in high dimensions ($d>2)$}
\label{sec:ap-high}

To determine the scaled mass density in $d$ dimensions, we insert the scaling ansatz  \eqref{scaling:rd} into the governing equation \eqref{cmrd} and deduce the scaling exponent $\beta=1-d/4$ together with the ODE for the scaled mass density
\begin{equation}
\label{Phi:d}
 \big(1-\tfrac{d}{4}\big)^{-1}(\Phi^2)'' + \big(1-\tfrac{d}{4}\big) x^2 \Phi''+ \big(3-\tfrac{d}{4}\big)x\Phi'+d\Phi = 0
\end{equation}
Multiplying by $x$ and integrating we obtain 
\begin{equation}
\label{Phi:2d}
\left(1-\frac{d}{4}\right)^{-2}\frac{d}{dx}\!\left(\frac{\Phi^2}{x}\right) + x\,\frac{d \Phi}{dx}+\frac{2d}{4-d}\,\Phi = 0
\end{equation}
Making the same transformation \eqref{Phi-Psi} and using again $Y=x^{3/2}$ we obtain
\begin{equation}
\label{implicit:Yd}
Y = \frac{6}{4-d}\big(C\,\psi^{-\delta} - \psi\big), \quad \delta = 3\,\frac{4-d}{4+3d}
\end{equation}
indicating that the results are applicable when $d<4$.

The same computation as in Eq.~\eqref{long-integral} allows one to fix the amplitude:
\begin{eqnarray*}
\frac{J}{(d-2)\Omega_d} & = &  \frac{2}{5}\int_0^{\psi_0} d\psi\, Y^\frac{5}{3} \\
                           & = & \frac{2}{5}\left(\frac{6}{4-d}\right)^\frac{5}{3}\int_0^{\psi_0} d\psi\, \big(C\,\psi^{-\delta} - \psi\big)^\frac{5}{3}\\
                           & = & \frac{2}{5}\left(\frac{6}{4-d}\right)^\frac{5}{3}\psi_0^\frac{8}{3}\int_0^1 du\, \big(u^{-\delta} - u\big)^\frac{5}{3}\\
                           & = & \frac{2}{5}\left(\frac{6}{4-d}\right)^\frac{5}{3}C^{\frac{4+3d}{6}}\,
                                    \frac{\Gamma(\Delta)\,\Gamma\!\left(\frac{8}{3}\right)}{(1+\delta)\,\Gamma\!\left(\Delta+\frac{8}{3}\right)}
\end{eqnarray*}
where $\Delta = (1+\delta)^{-1} - 5\delta/3$. Thus 
\begin{equation}
C = (A_dJ)^{6/(4+3d)}
\end{equation}
with a cumbersome expression for the numerical factor $A_d$ following from above formulas.


\begin{thebibliography}{99}

\bibitem{meakin}
       P.~Meakin, {\it Fractals, Scaling and Growth Far From Equilibrium}
       (Cambridge University Press, New York, 1998).

\bibitem{az}
       A.~Zangwill, {\it Physics at Surfaces}
       (Cambridge University Press, New York, 1988).

\bibitem{ls}
       I.~M.~Lifshitz and V.~V.~Slyozov,
       Zh.\ Eksp.\ Teor.\ Fiz. {\bf 35}, 479 (1959)
       [Sov.\ Phys.\ JETP {\bf 8}, 331 (1959)]; J. Phys.\ Chem.\ Solids
       {\bf 19}, 35 (1961).

\bibitem{ajb}
       A.~J.~Bray, Adv. Phys. {\bf 43}, 357 (1994).

\bibitem{sm}
       C.~Sire and S.~N.~Majumdar,
       Phys.\ Rev.\ E {\bf 52}, 244 (1995).

\bibitem{ts}
       T.~Schelling, J.\ Math.\ Sociology {\bf 1}, 61 (1971).

\bibitem{Angle86}
       J. Angle, Social Forces {\bf 65}, 293 (1986).
        
\bibitem{IKR98}
       S.~Ispolatov, P.~L.~Krapivsky, and S.~Redner,
       Eur.\ Phys.\ J. B {\bf 2}, 267 (1998).

\bibitem{CC00}
      A. Chakraborti and  B. K. Chakrabarti, Eur. Phys. J. B {\bf 17}, 167 (2000). 

\bibitem{Hayes}
      B. Hayes, American Scientist {\bf 90}, 400 (2002).
      
\bibitem{Angle06}
     J. Angle, Physica A {\bf 367}, 388 (2006). 

\bibitem{Boghosian15}
     B. M. Boghosian, M. Johnson, and J. A. Marcq, J. Stat. Phys. {\bf 161}, 1339 (2015).

\bibitem{lr}
       F. Leyvraz and S. Redner,
       Phys.\ Rev.\ Lett. {\bf 88}, 068301 (2002).

\bibitem{Sun14}
       R. Sun, Physics Letters A {\bf 378}, 3177 (2014).        

\bibitem{GG09} 
          P. Gaspard and T. Gilbert, J. Stat. Mech. P08020 (2009). 

\bibitem{Khanin12}
          A. Grigo A, K. Khanin, and D. Sz\'asz,  Nonlinearity {\bf 25},  2349 (2012).
          
\bibitem{Evans12}
       M. R. Evans and B. Waclaw,  Phys. Rev. Lett. {\bf 108}, 070601 (2012). 

\bibitem{Colm15}
       Yu-Xi Chau, C. Connaughton, and S. Grosskinsky, J. Stat. Mech. P11031 (2015). 

\bibitem{GG17} 
          P. Gaspard and T. Gilbert, J. Stat. Mech. P043210 (2017). 

\bibitem{KL02}
       J.~Ke and Z.~Lin,
       Phys. Rev. E {\bf 66}, 050102 (2002).

\bibitem{EP03}
      E.~Ben-Naim and P.~L.~Krapivsky, Phys.\ Rev.\ E {\bf 68}, 031104 (2003).

\bibitem{KL06}
       Z.~Lin, J.~Ke, and G.~Ye, 
       Phys. Rev. E {\bf 74}, 046113 (2006).

\bibitem{book}  
      P. L. Krapivsky, S. Redner and E. Ben-Naim,
      {\it  A Kinetic View of Statistical Physics}
      (Cambridge: Cambridge University Press, 2010).

\bibitem{PK_exchange}
       P.~L.~Krapivsky,    J. Phys. A  {\bf  48}, 205003 (2015).

\bibitem{ZK} 
     Ya. B. Zeldovich and A. S. Kompaneets, in:
     {\it Collection of Papers Dedicated to A. F. Ioffe}, pp. 61--71
     (Izd. Akad. Nauk USSR, Moscow, 1950).     

\bibitem{B52} 
     G.~I.~Barenblatt, Prikl. Mat. Mekh. {\bf 16}, 67 (1952).

\bibitem{Zeld} 
     Ya. B. Zeldovich and Yu. P. Raizer, 
     {\it Physics of Shock Waves and High-Temperature Hydrodynamic Phenomena}, Vol. I \& II
     (Academic Press, New York, 1966).     

\bibitem{LL87} 
     L. D. Landau and E. M. Lifshitz, {\it Fluid Mechanics}
    (Pergamon Press, New York, 1987).

\bibitem{B96} 
     G.~I.~Barenblatt, {\it Scaling, Self-Similarity, and Intermediate
     Asymptotics} (Cambridge University Press, Cambridge, 1996).

\bibitem{Landim:09} 
     P. Gon\c{c}alves, C. Landim, and C. Toninelli, Ann. l'Inst. H. Poincar\'{e} -- Probab. Statist. {\bf 45}, 887 (2009). 

\bibitem{Pablo12}
     P.~I.~Hurtado and P.~L.~Krapivsky,  Phys.\ Rev.\ E {\bf 85}, 060103(R) (2012). 

\bibitem{vD89}
     P. G. J. van Dongen, Phys. Rev. Lett. {\bf 63}, 1281 (1989).
     
\bibitem{S88} 
     J.~L.~Spouge, Phys.\ Rev.\ Lett.\ {\bf 60}, 871 (1988).

\bibitem{TNT88}
      H. Takayasu,  I. Nishikawa, and H. Tasaki, Phys.\ Rev.\  A {\bf 37}, 3110 (1988).

\bibitem{T89}
     B.~R.~Thomson, J. Phys.\ A {\bf 22}, 879 (1989).

\bibitem{Andrews}
     G. E. Andrews,  {\em The Theory of Partitions} (Cambridge University Press, New York, 1976).

\bibitem{Vershik}   
     A. M. Vershik, Funct. Anal. Appl. {\bf 30}, 90 (1996);  
     A. M. Vershik, J. Math. Sci. {\bf 119}, 165 (2004).

\bibitem{Euler:book}
     L. Euler, {\em Introduction to Analysis of the Infinite} (Springer, New York, 1988).

\bibitem{BD99}
    M. Bauer and D. Bernard, J. Phys. A {\bf 32}, 5179 (1999).

\bibitem{E00}
    W. E and E. Vanden-Eijnden, Commun. Pure Appl. Math. {\bf 53}, 852 (2000).

\bibitem{Khanin07}
     J. Bec and K. Khanin,  Phys. Rep. {\bf 447}, 1 (2007).

\bibitem{Anna14}
     A. Frishman and G. Falkovich, Phys. Rev. Lett. {\bf 113}, 024501 (2014). 
     
\bibitem{EP}
      D. Landolt,  Electrochimica Acta {\bf 32}, 1 (1987).
      
\bibitem{D87}
      G. Daccord, Phys. Rev. Lett. {\bf 58}, 479 (1987).

\bibitem{KM91}    
       J. Krug and P. Meakin, Phys. Rev. Lett. {\bf 66}, 703 (1991).

\bibitem{SBG} 
      B. Sapoval, A. Baldassarri, and A. Gabrielli, Phys. Rev. Lett. {\bf 93}, 098501 (2004).

\bibitem{Lar}    
       H.~Larralde, Y. Lereah, P. Trunfio, J. Dror, S. Havlin, R. Rosenbaum, 
       and H. E. Stanley, Phys.\ Rev.\ Lett. {\bf 70}, 1461 (1993).   

\bibitem{PLK:Stefan}     
     P. L. Krapivsky, Phys. Rev. E {\bf 85}, 031124 (2012). 

\bibitem{IDLA_1}  
      A. Asselah and A. Gaudilli\`ere, Ann. Probab. {\bf 41}, 1160 (2013). 

\bibitem{IDLA_2}    
      D. Jerison, L. Levine, and S. Sheffield, J. Am. Math. Soc. {\bf 25}, 271 (2012);
      Duke Math. J. {\bf 163}, 267 (2014). 
      
\bibitem{Sid89}     
     Z. Cheng, S. Redner, and F. Leyvraz, Phys. Rev. Lett. {\bf 62}, 2321 (1989).

\bibitem{PLK:3particle}     
      P. L. Krapivsky, Phys. Rev. E {\bf 49}, 3233 (1994).

\bibitem{PLK:source}     
      P. L. Krapivsky, Physica A {\bf 198}, 157 (1993).

\bibitem{Asymmetric:source}
      H. Hinrichsen, V. Rittenberg, and H. Simon, J. Stat. Phys. {\bf 86}, 1203 (1997). 

\bibitem{Kirone:source}     
      A. Ayyer and K. Mallick,  J. Phys. A: Math. Theor. {\bf 43}, 045003 (2010).  

\bibitem{PLK12}     
      P. L. Krapivsky, Phys. Rev. E {\bf 86}, 041113 (2012). 

\bibitem{einstein}  
      S.~V.~Khare, N.~C.~Bartelt, and T.~L.~Einstein, Phys.\ Rev.\ Lett. {\bf 75}, 2148 (1995).    
  
\bibitem{BK91}
       N.~V.~Brilliantov and P.~L.~Krapivsky,
       J. Phys.\ A: Math. Gen. {\bf 24}, 4789 (1991).
       
\bibitem {Jose98}
      P.~L.~Krapivsky, J.~F.~F.~Mendes, and S.~Redner,   Eur.\ Phys.\ J.\ B {\bf 4}, 401 (1998). 

\bibitem {Jose99}
      P.~L.~Krapivsky, J.~F.~F.~Mendes, and S.~Redner,  Phys.\ Rev.\ B {\bf 59}, 15950 (1999). 

       
\end{thebibliography}
\end{document}